\documentclass{article}

\usepackage{arxiv}
\usepackage{array}
\usepackage[utf8]{inputenc} 
\usepackage[T1]{fontenc}    
\usepackage{hyperref}       
\usepackage{url}            
\usepackage{booktabs}       
\usepackage{amsfonts}       
\usepackage{nicefrac}       
\usepackage{microtype}      
\usepackage{lipsum}		
\usepackage{graphicx}
\graphicspath{ {./fig/} }
\usepackage[square,numbers]{natbib}
\usepackage{multirow}
\usepackage{doi}
\usepackage{nomencl}
\usepackage[ruled]{algorithm2e}
\usepackage{float}
\usepackage{enumitem}
\usepackage{pifont}

\usepackage[most]{tcolorbox}
\usepackage{xcolor}
\usepackage{lmodern}

\usepackage{listings}
\usepackage{xcolor}
\usepackage[most]{tcolorbox}

\usepackage{caption}

\usepackage{tabularx}

\tcbset{
  colback=gray!10,     
  colframe=gray!30,    
  boxrule=1pt,         
  arc=1mm,             
  left=2mm, right=2mm, top=1mm, bottom=1mm,
  fontupper=\ttfamily, 
  enhanced
}

\title{Intelligent Design 4.0: \\Paradigm Evolution Toward the Agentic AI Era}
\date{A Preprint Version: Oct 20, 2025}

\author{ 
{
\hspace{1mm}Shuo Jiang\thanks{Comments are welcome: \texttt{shuojiangcn@gmail.com}}} \\
	Department of Systems Engineering \\
        City University of Hong Kong, Hong Kong \\
	\texttt{shuo.jiang@cityu.edu.hk} \\
	\And
        {
        {\hspace{1mm}Min Xie}} \\
	Department of Systems Engineering \\
        City University of Hong Kong, Hong Kong \\
	\texttt{minxie@cityu.edu.hk} \\
        \And
        {
        {\hspace{1mm}Frank Youhua Chen}} \\
	Department of Decision Analytics and Operations \\
        City University of Hong Kong, Hong Kong \\
	\texttt{youhchen@cityu.edu.hk} \\
        \And
        {
        {\hspace{1mm}Jian Ma}} \\
	Department of Information Systems \\
        City University of Hong Kong, Hong Kong \\
	\texttt{isjian@cityu.edu.hk} \\
        \And
        {
	{\hspace{1mm}Jianxi Luo$^{*}$}} \\
	Department of Systems Engineering \\
        City University of Hong Kong, Hong Kong \\
	\texttt{jianxi.luo@cityu.edu.hk} \\
}

\renewcommand{\headeright}{Intelligent Design 4.0: Paradigm Evolution Toward the Agentic AI Era}
\renewcommand{\undertitle}{}
\renewcommand{\shorttitle}{}

\begin{document}
\maketitle

\begin{abstract}
	Research and practice in Intelligent Design (ID) have significantly enhanced engineering innovation, efficiency, quality, and productivity over recent decades, fundamentally reshaping how engineering designers think, behave, and interact with design processes. The recent emergence of Foundation Models (FMs), particularly Large Language Models (LLMs), has demonstrated general knowledge-based reasoning capabilities, and open new avenues for further transformation in engineering design. In this context, this paper introduces \textbf{Intelligent Design 4.0 (ID 4.0)} as an emerging paradigm empowered by foundation model-based agentic AI systems. We review the historical evolution of ID across four distinct stages: rule-based expert systems, task-specific machine learning models, large-scale foundation AI models, and the recent emerging paradigm of foundation model-based multi-agent collaboration. We propose an ontological framework for ID 4.0 and discuss its potential to support end-to-end automation of engineering design processes through coordinated, autonomous multi-agent-based systems. Furthermore, we discuss challenges and opportunities of ID 4.0, including perspectives on data foundations, agent collaboration mechanisms, and the formulation of design problems and objectives. In sum, these insights provide a foundation for advancing Intelligent Design toward greater adaptivity, autonomy, and effectiveness in addressing the growing complexity of engineering design.
\end{abstract}


\newpage

\section{Introduction}
\label{sec1}

Design is a central activity in engineering \cite{Dym1994,Ulrich2016}. It serves as the engine for transforming multi-level requirements and constraints into functional, manufacturable, and impactful products or solutions \cite{Beitz1996}. Historically grounded in expert heuristics and iterative human evaluation, design methodology has evolved with the development of computational tools, including drafting and modeling software such as Computer-Aided Design (CAD), simulation and optimization toolkits in Computer-Aided Engineering (CAE), and ideation support tools in Computer-Aided Ideation (CAI). As design problems become more complex, interdisciplinary, and dynamic, traditional design paradigms face limitations in terms of scalability and adaptivity.

Intelligent Design (ID) has emerged as a cross-disciplinary field at the intersection of engineering and artificial intelligence, with the goal of embedding reasoning, learning, and automation into design processes. ID includes a broad set of methods, tools, and systems that augment or automate design tasks through artificial intelligence. ID shares conceptual and methodological overlap with terms such as Design Automation \cite{Bliek1992} and Data-Driven Design \cite{Kim2016}, while covering a more inclusive scope that unifies all forms of computational intelligence aimed at supporting the engineering design process. Over successive technological waves, from rule-based or knowledge-based engineering systems \cite{Chapman1999}, data-driven inference \cite{Luo2022}, to deep generative models \cite{Regenwetter2022review}, ID has reshaped how design is represented, explored, and realized.

More recently, the emergence of large-scale foundation models (e.g., Large Language Models (LLMs) and Vision Language Models (VLMs)) pretrained on vast corpora of multimodal data and refined through supervised fine-tuning and reinforcement learning, has demonstrated broad general-purpose capabilities \cite{OpenAI2025,Anthropic2025,Meta2025}. These models have shown promise across diverse domains, including scientific discovery and engineering practice \cite{Boiko2023,Romera-Paredes2024}. In engineering design, recent research has explored their application in design tasks such as requirement extraction \cite{Ataei2025}, design ideation \cite{Jiang2025autotriz}, design modeling \cite{Li2025}, and design optimization \cite{Jiang2025dsm}. However, these efforts largely support only one or several isolated stages of the overall design process and continue to rely heavily on human efforts for coordination and decision-making, particularly for system-level engineering design.

Building upon existing foundation models, recent advances point toward the emergence of AI agents \cite{Durante2024,Tian2025}. These AI agent-based systems that not only leverage the reasoning capacity of foundation models but also incorporate memory, interactive action spaces with external tools and environments, and the ability to continuously learn from experience. Compared with the early agent-based systems of the 1990s to 2010s \cite{SoriaZurita2018,Panchal2009,Tsompanopoulou2008}, which were mainly rule-based or process-driven, while LLM-based agents provide qualitatively new capabilities. For example, they can perform general-purpose reasoning across domains, maintain contextual memory for adaptive learning, and autonomously interact with external tools such as web search, databases, and other agents. By combining the complementary strengths of multiple different agents, the systems can accomplish higher-level and more complex tasks \cite{Sun2025}. Multi-agent systems have already been explored in other fields \cite{Moritz2025,Gao2024,Swanson2025}. In healthcare, MASH system integrates multiple specialist agents into a coordinated network, outperforming single-model approaches in multi-turn clinical tasks \cite{Moritz2025}. In biomedical discovery, AI agents serve as autonomous “AI scientists,” orchestrating hypothesis generation, experimental planning, and iterative refinement across disciplines \cite{Gao2024}. Most recently, a multi-agent virtual laboratory successfully designed novel SARS-CoV-2 nanobodies, two of which demonstrated improved binding to emerging variants \cite{Swanson2025}. In the field of intelligent design, such multi-agent-based systems are particularly powerful because design tasks span multiple interdependent stages that cannot be effectively handled by a single model. For example, a recent exploratory study integrated the expertise of AI agents across mechanical design, optimization, electronics, and software engineering, enabling the autonomous generation of functional prototypes with minimal direct human input, and ultimately producing an autonomous vessel \cite{Wang2025agent}. Multi-agent collaboration can distribute these tasks among specialized agents, enable iterative exchange of intermediate representations, and dynamically coordinate across CAD/CAE workflows. These capabilities suggest that agentic systems will fundamentally reshape the design process by reducing reliance on human expert and enabling more autonomous and end-to-end innovation.

Herein, this paper positions a new paradigm termed Intelligent Design 4.0 (ID 4.0). We propose a unifying conceptual and ontological framework that organizes emerging developments in agentic AI into a coherent design paradigm. While some domain-specific demonstrations of agent collaboration have appeared, the field still lacks a systematic, model-based understanding of how such agents can be architected, coordinated, and scaled across the entire design process. ID 4.0 builds this foundation through a design process-driven ontology that connects design data, agent collaboration, and problem formulation, enabling a pathway toward end-to-end automation in engineering design. In the following sections, we retrospectively analyze the evolution of Intelligent Design across four developmental stages, present the conceptual framework of ID 4.0, and discuss key challenges and opportunities for advancing this paradigm.

\section{The Evolution of Intelligent Design Paradigms}
\label{sec:sec2}

Over the past decades, Intelligent Design has evolved from static, manually encoded expert systems to increasingly adaptive, data-driven, and now potentially autonomous systems powered by multiple AI agents. This paradigm evolution is mainly driven by three essential shifts:

\begin{enumerate}[label=(\arabic*)]
  \item The expansion of the design-related knowledge of AI, from limited and domain-specific datasets to vast and cross-domain data sources.
  \item The progression of AI capabilities from task-specific assistance to general-purpose reasoning and decision-making.
  \item The transition from human-in-the-loop control toward partially or fully autonomous design agents.
\end{enumerate}

These shifts are redefining how design problems are framed, how solution spaces are explored, and how final design decisions are made. We identify four major stages in the trajectory of ID’s evolution, as illustrated in Figure 1.

\begin{figure}[H]
	\centering
	\includegraphics[width=16cm]{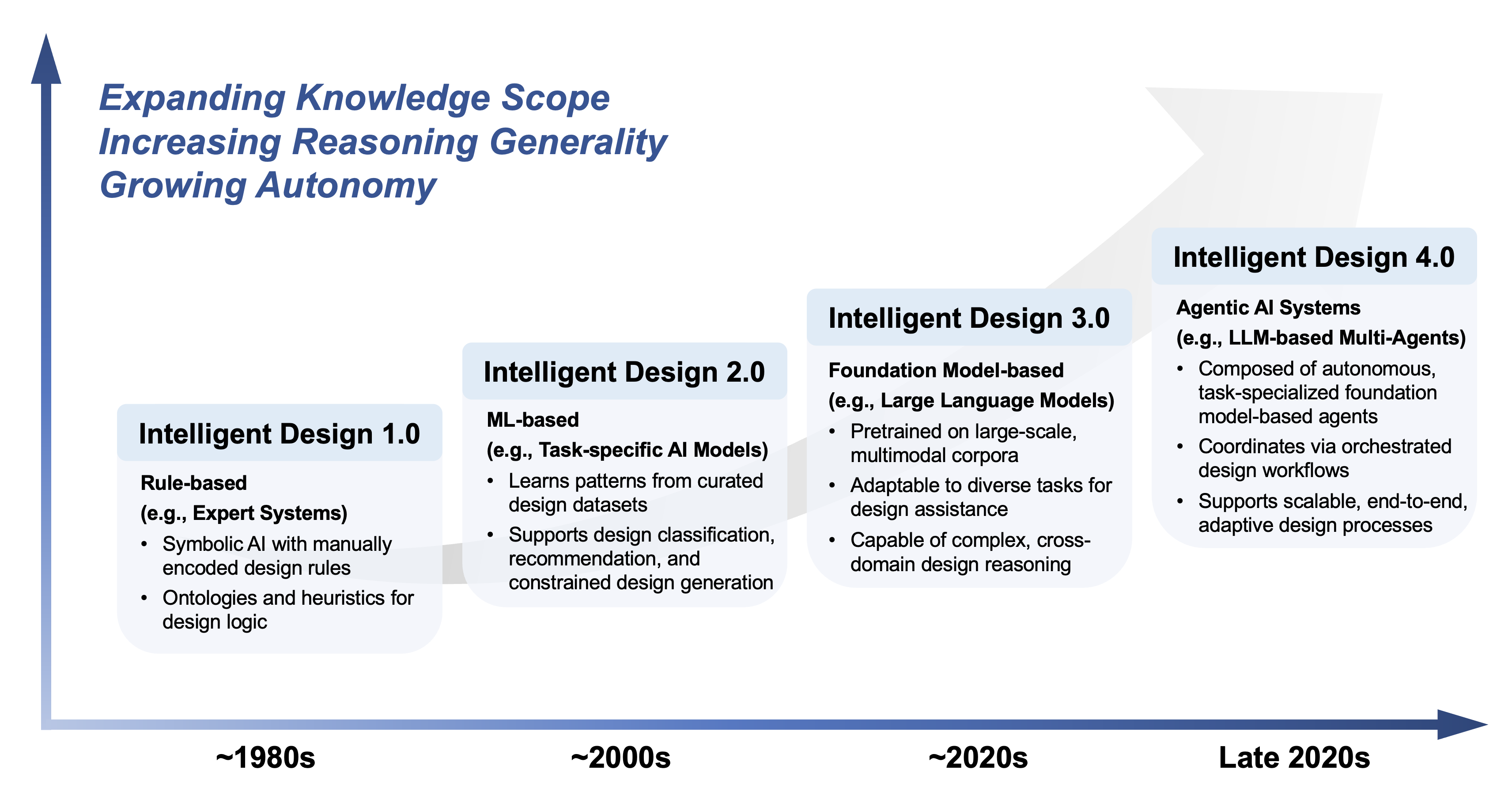}
	\caption{The Four-stage Evolution of the Intelligent Design Paradigm}
	\label{fig:fig1}
\end{figure}

\section{Intelligent Design 1.0: Rule-Based Expert Systems}
\label{sec3}

The first generation of Intelligent Design systems emerged in the 1980s-1990s, grounded in symbolic AI and expert systems. These rule-based systems aimed to capture and formalize expert knowledge into machine-readable formats for reuse. The underlying assumption was that design knowledge could be articulated explicitly and embedded into deterministic and heuristic rules \cite{Brown1986}.

A foundational aspect of ID 1.0 was the development of design ontologies and representation frameworks, such as the Function-Behavior-Structure (FBS) \cite{Qian1996}, SAPPhIRE \cite{Venkataraman2009}, Functional basis \cite{Hirtz2002}, and Environment-Based Design (EBD) \cite{Zeng2004} ontologies. These frameworks enabled structured reasoning about design intent, function realization, and system behavior, facilitating design knowledge reuse \cite{Dym2012}. In terms of methodology, ID 1.0 systems were often built upon structured design methods such as TRIZ, C-K Theory, Quality Function Deployment (QFD), and Axiomatic Design, which provided prescriptive strategies for exploring and evaluating solution spaces \cite{Hatchuel2003,Altshuller1999,Suh1998,Akao2024,Tan2003}. In parallel, early CAD software systems (e.g., AutoCAD, SolidWorks, Pro/E) also played a foundational role in Intelligent Design by automating repetitive operations and modeling tasks. While not traditionally categorized as AI, these tools significantly reduced manual effort and laid the groundwork for embedding structured logic into design workflows.

The computational realization of ID 1.0 was exemplified in early Knowledge-Based Engineering (KBE) and Case-Based Reasoning (CBR) systems, such as KRITIK \cite{Goel2014}, Argo \cite{Huhns1988}, DANE \cite{Vattam2011}, and InnoGPS \cite{Luo2021}, which supported analogy-based design through knowledge retrieval and re-organization. In addition, Design Structure Matrix methods and relevant tools were also widely used to represent and manage interdependencies in complex engineering systems \cite{Eppinger2012}. While effective in specific domains, these systems were inherently static, limited by their reliance on manually curated knowledge and rule sets. Figure 2 presents representative ID 1.0 systems that demonstrate early efforts in rule-based and knowledge-based design support.

\begin{figure}[H]
	\centering
	\includegraphics[width=16cm]{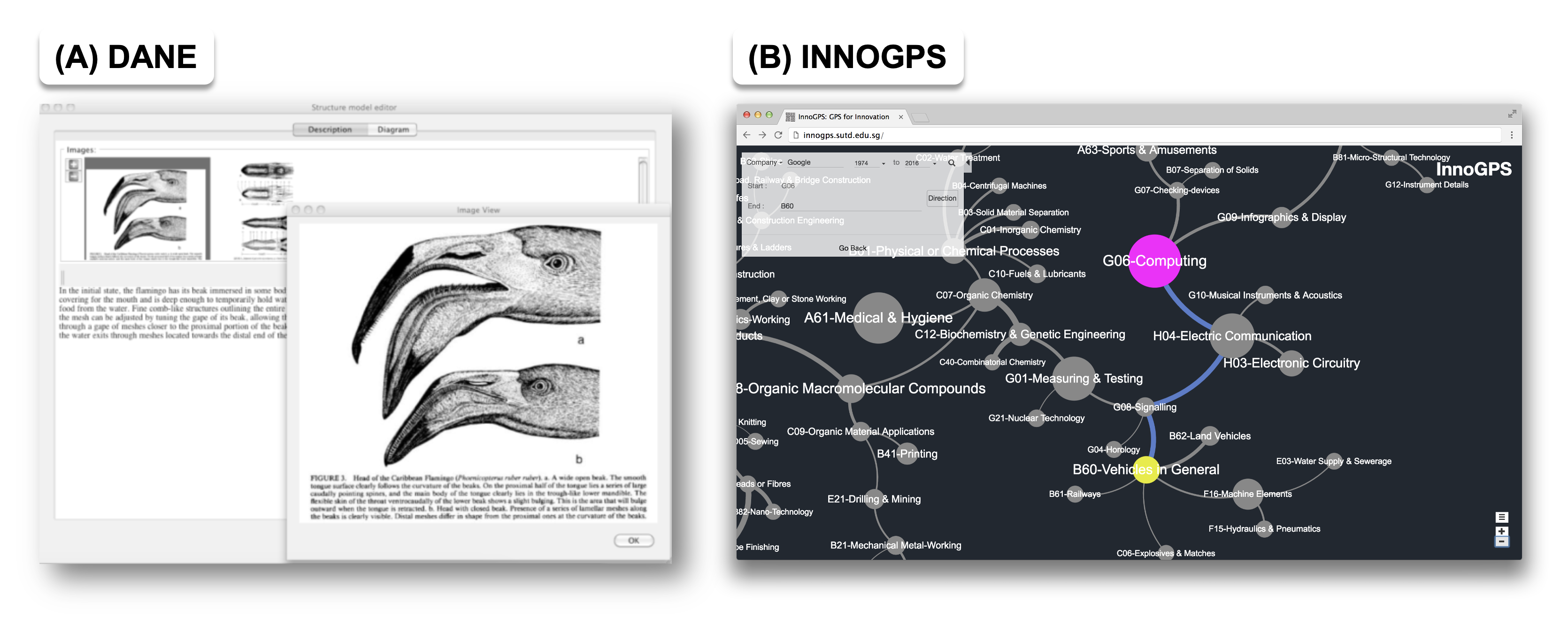}
	\caption{Representative ID 1.0 Systems: (A) DANE \cite{Vattam2011}, a design by analogy to nature engine; (B) INNOGPS \cite{Luo2021}, a computer-aided system for design ideation and exploration}
	\label{fig:fig2}
\end{figure}

\section{Intelligent Design 2.0: Machine Learning-Driven Design}
\label{sec:sec4}

With the rise of computational power and the advancement of statistical learning techniques, especially the deep neural networks, ID 2.0 emerged with the widespread adoption of machine learning (ML) and deep learning (DL) methods \cite{LeCun2015}. ID 2.0 systems can learn correlations and latent structures from small-to-medium-scale datasets curated to specific design tasks, presenting a shift from expert-derived rules to data-driven computational reasoning.

One major direction of ID 2.0 was the development of systems for design information retrieval and design inspiration recommendation. Initial work leveraged structured textual data, such as thesaurus-based retrieval of informal design descriptions \cite{Yang2005}. Later research extended these capabilities to include technical documents \cite{Murphy2014}, images \cite{Jiang2021}, sketches \cite{Zhang2022}, and multimodal datasets that integrated textual, visual, and 3D models to enhance the discovery of analogous or novel design stimuli \cite{Kwon2022}.

For design method recommendation, ML and DL algorithms were employed to identify relevant design principles or process heuristics. For example, supervised models have been trained to recommend suitable methods based on problem representations and user intent \cite{Fuge2014}. Another study mined latent patterns in designers’ workflows using Hidden Markov Models (HMM), which captured the sequential structure of human decision-making in design tasks for new design generation \cite{McComb2017}.

In recent years, Generative AI (GenAI) models introduced a new form of machine creativity \cite{Regenwetter2022review}. Variational Autoencoders (VAEs), Generative Adversarial Networks (GANs), and their derivatives were developed and adopted to generate novel design candidates under functional, structural, and performance constraints \cite{Regenwetter2022review}. These models have been applied on various design cases, such as wheels \cite{Oh2019}, bicycles \cite{Regenwetter2022biked}, airfoils \cite{Chen2019}, and vehicles \cite{Chilukuri2024}, demonstrating the power of GenAI methods in constrained design synthesis.

Additionally, techniques such as Natural Language Processing (NLP) \cite{Siddharth2022}, Semantic Networks (SN) \cite{Han2022}, and Multi-Modal Machine Learning \cite{Song2024} also expanded the ability to model and reason over complex design-related tasks. By leveraging various forms of structured and unstructured data \cite{Luo2022,Jiang2022}, ID 2.0 systems enabled more intelligent representation, search, recommendation, and generation, ultimately improving design efficiency, creativity, and decision support across a wide range of tasks. Figure 3 presents representative ID 2.0 systems, which leverage ML/DL techniques to support specific tasks in the engineering design workflow.

\begin{figure}[H]
	\centering
	\includegraphics[width=16cm]{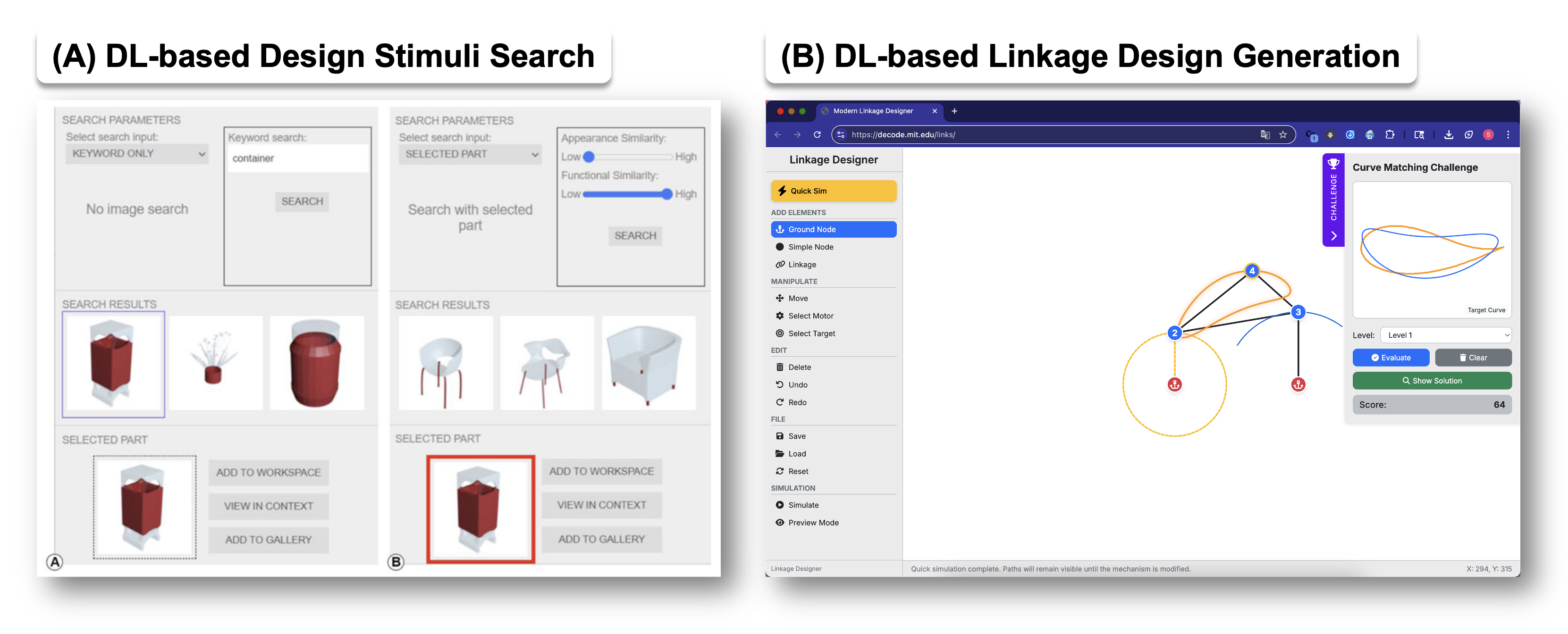}
	\caption{Representative ID 2.0 Systems: (A) DL-based design stimuli search system that retrieves analogous visual designs \cite{Kwon2022}; (B) DL-based linkage design generation system that synthesizes mechanical linkages \cite{Nobari2024}}
	\label{fig:fig3}
\end{figure}

\section{Intelligent Design 3.0: Foundation Model-Empowered Design}
\label{sec:sec5}

The third stage in the evolution of Intelligent Design is driven by the introduction of large-scale foundation AI models into engineering design workflows, including LLMs and VLMs. These models are pretrained on vast multimodal corpora and exhibit emergent abilities such as in-context learning, instruction following, and step-by-step reasoning \cite{Zhao2023,Chang2024}. In engineering design, recent studies have systematically evaluated their applicability and limitations on design-related tasks, demonstrating significant potential to support various stages of the design process \cite{Makatura2024,Picard2023,Ma2025,Zhu2023bio,Zhu2023gpt}.

Despite their remarkable reasoning power and adaptability, foundation AI models face critical challenges, such as hallucination, reasoning inconsistency, and the lack of verifiability \cite{Ji2023}. To mitigate these limitations, researchers have proposed new methodologies aimed at better aligning model behavior with design-specific requirements. These include:
\begin{enumerate}[label=(\arabic*)]
  \item defining explicit task workflows to guide the reasoning steps of LLMs in requirement analysis, design generation, and evaluation;
  \item combining lightweight task-specific models with general-purpose LLMs to improve output fidelity and domain relevance;
  \item integrating retrieval-augmented generation (RAG) or domain-specific fine-tuning to reinforce factual grounding and context consistency.
\end{enumerate}

With these strategies, researchers have developed LLM-based and VLM-based approaches to enhance one or multiple stages of the design processes, including requirement analysis \cite{Ataei2025,Fabunmi2025,Zhu2025empathy}, concept generation \cite{Jiang2025autotriz,Zhu2023bio,Chen2025,Ren2025}, system construction \cite{Gomez2024}, CAD modeling and editing \cite{Li2025,Zhang2025cad,GregSweeneyZooCorporation2025}, and design optimization \cite{Jiang2025dsm,Zhang2025autoturb}. For example, in the area of concept generation, we proposed AutoTRIZ \cite{Jiang2025autotriz}, which integrates the reasoning capacity and broad domain knowledge of LLMs with the TRIZ-based problem-solving framework. In AutoTRIZ, the reasoning process of the LLM is controlled to strictly follow TRIZ logic, while an external knowledge base provides stage-specific TRIZ-related knowledge, thereby improving the accuracy and relevance of generated solutions. We applied AutoTRIZ to the design of a battery thermal management system (BTMS) and successfully generated solutions that balanced both thermal efficiency and system performance compared with previous work. In the area of optimization, we applied LLMs to the optimization of DSMs by combining the topological and semantic information in DSMs \cite{Jiang2025dsm}. Experimental results demonstrated that the proposed LLM-based optimization method achieved faster convergence and superior solution quality compared to benchmark approaches. Moreover, we found incorporating contextual domain knowledge of DSMs further enhanced optimization performance. These results highlight that the semantic understanding and complex reasoning abilities of foundation models hold broad potential for addressing real-world engineering design problems.

The above efforts reflect a shift from using foundation AI models as general copilots toward positioning them as adaptable, architecture-integrated components within intelligent design systems. Compared to ID 2.0, where models mainly automated decisions or learned patterns from labeled data, ID 3.0 systems are increasingly capable of complex reasoning and reflective decision-making. Figure 4 presents representative ID 3.0 systems that leverage large-scale foundation models for conceptual ideation and detailed CAD modeling.

\begin{figure}[H]
	\centering
	\includegraphics[width=16cm]{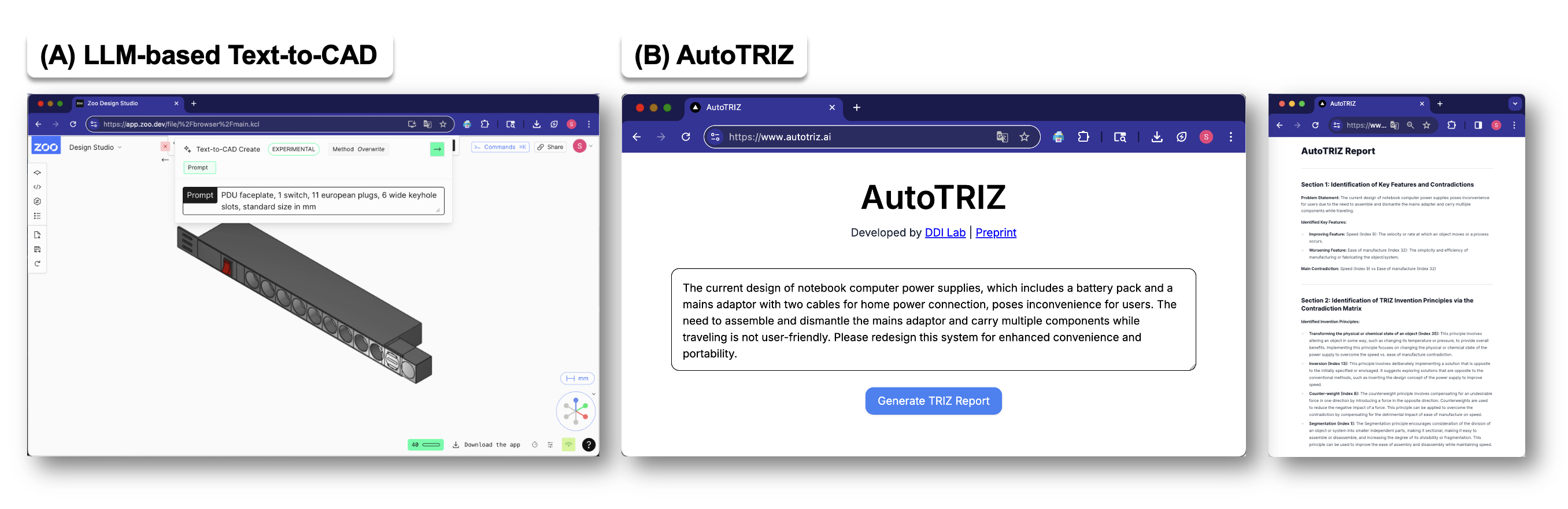}
	\caption{Representative ID 3.0 Systems: (A) ZOO \cite{GregSweeneyZooCorporation2025}, an LLM-based Text-to-CAD system; (B) AutoTRIZ \cite{Jiang2025autotriz}, an artificial ideation system with TRIZ and LLMs}
	\label{fig:fig4}
\end{figure}

\section{Intelligent Design 4.0: Foundation Model-based Multi-Agent AI Systems}
\label{sec:sec6}

While LLM-based methods present advanced capabilities to design, a single foundation model remains insufficient to handle the complexity of the full stack of engineering design. The entire design process spans a wide range of cognitive, computational, and domain-specific subtasks. The design methods and tools of ID 3.0 still require extensive human efforts for task decomposition and coordination \cite{Wang2025custom}. In addition, recent studies have also revealed intrinsic capability variations across different models \cite{Hu2024}. For instance, some LLMs such as Claude models excel at code generation, while others like GPT models demonstrate stronger performance in text generation \cite{OpenAI2025,Anthropic2025}. These observations suggest that the complex design tasks, are better addressed not by a single foundation model, but by an ensemble of specialized, collaborative AI agents, each fine-tuned or prompted for particular functions \cite{Durante2024}.

\begin{figure}[H]
	\centering
	\includegraphics[width=15cm]{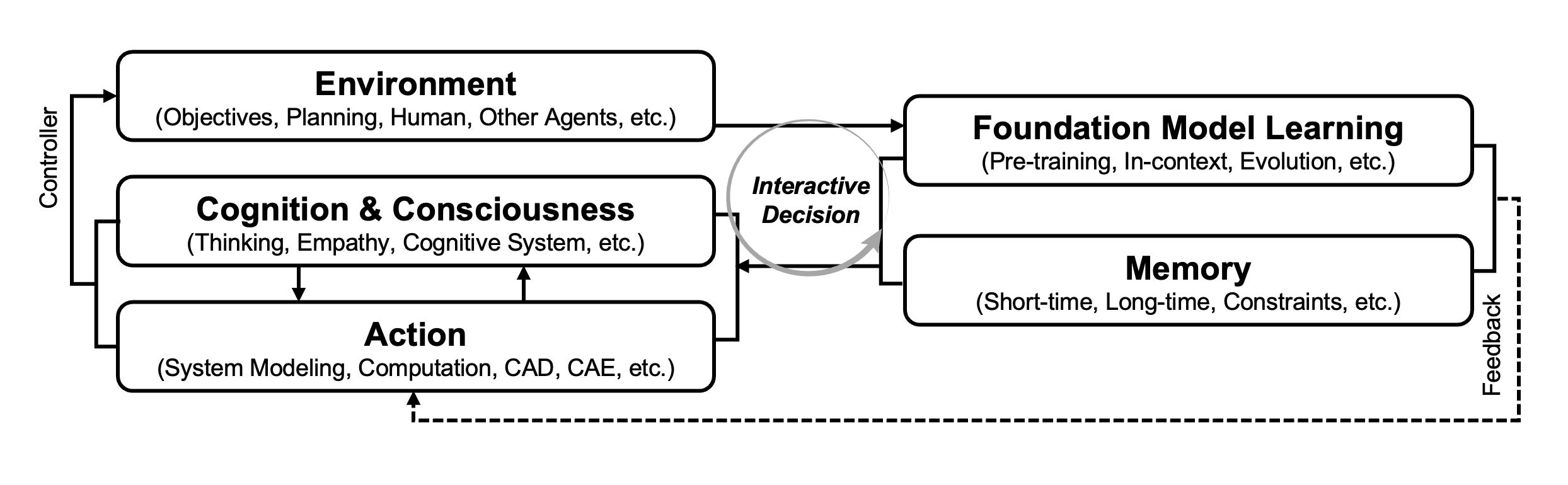}
	\caption{The Framework of an Exemplar AI Agent}
	\label{fig:fig5}
\end{figure}

The idea of agent-based modeling was first proposed in the late 1940s but became widely used only in the 1990s because of its high computational demands \cite{JohnVonNeumann1966}. Early agent-based modeling is a computational approach that simulates the actions and interactions of autonomous agents (individuals, groups, or organizations) to study system behaviors and the factors that shape their outcomes. Different from early agent-based systems that relied on pre-defined rules or equation-based simulations \cite{SoriaZurita2018,Panchal2009,Tsompanopoulou2008}, foundation model-driven agents can learn from diverse data, adapt to new contexts, and generate flexible strategies beyond hand-crafted assumptions. Figure 5 illustrates the architecture of an exemplar AI agent, consisting of environment, cognition, action, memory, and foundation model learning modules. The framework highlights how interactive decision-making connects these components, enabling continuous adaptation and feedback.

Building upon the foundation of AI agents, multi-agent systems can collaborate and coordinate with one another to accomplish more complex system-level objectives. Multi-agent systems have been explored and applied across diverse domains such as biomedicine \cite{Gao2024}, healthcare \cite{Moritz2025}, scientific discovery \cite{Ghafarollahi2025}, and software development \cite{Qian2023}. In the field of engineering design, some recent studies also explored leveraging multiple agents to conduct complex design tasks \cite{Massoudi2025,Panta2025}. For example, Massoudi and Fuge \cite{Massoudi2025} evaluated a structured multi-agent system for early-stage engineering design and showed that multi-agent orchestration, compared with a simpler two-agent setup, can generate more detailed design-state representations and improve workflow management. Panta et al. \cite{Panta2025} proposed Mechanical Engineering Design Agents (MEDA), a multi-agent framework that leverages multimodal LLMs for automated parametric CAD modeling, demonstrating its effectiveness in achieving near-perfect script execution and significantly improving model accuracy on a CAD dataset. In Figure 6, we synthesize these insights and propose a conceptual framework for implementing ID 4.0. The end-to-end, design-process-based ontological framework can serve to guide the growing but scattered efforts toward developing multi-agent systems for design automation.

\begin{figure}[H]
	\centering
	\includegraphics[width=16cm]{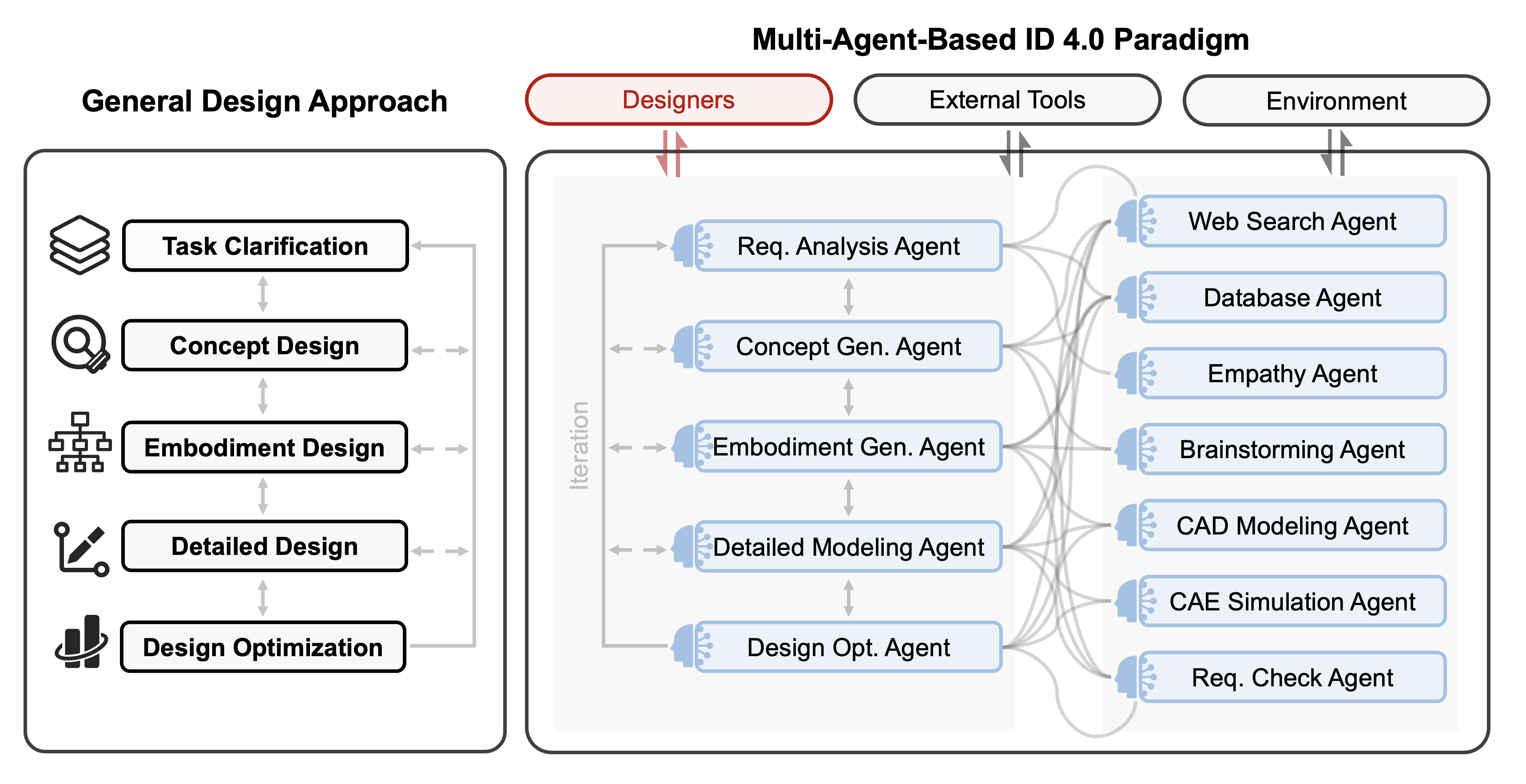}
	\caption{An Ontological Framework for ID 4.0: Multi-Agent-Based End-to-End Design Automation}
	\label{fig:fig6}
\end{figure}

The left panel of Figure 6 outlines a five-stage engineering design workflow adapted from \cite{Beitz1996}: Task Clarification, Concept Design, Embodiment Design, Detailed Design, and Design Optimization. This framework is widely adopted in engineering design, with its stages forming an iterative loop in which solutions are progressively refined and evaluated. The right panel maps this workflow onto a multi-agent AI system, where each stage is orchestrated by a corresponding stage-level agent:

\textbf{Requirement Analysis Agent (RAA)} interprets user inputs (typically multimodal information such as text or images), extracts design goals (e.g., functional and performance requirements), identifies constraints (e.g., usage environment, manufacturing conditions, cost limitations), and decomposes high-level tasks into structured sub-problems for downstream agents as needed. Beyond simple interpretation, the RAA also ensures traceability between initial requirements and subsequent design stages, thereby anchoring the overall workflow. In engineering contexts, this traceability must bridge heterogeneous data sources, linking textual requirements with quantitative design parameters, simulation constraints, and manufacturing considerations. This remains a persistent challenge in model-based systems engineering. The RAA thus serves as an interface between unstructured human intent and structured engineering models.

\textbf{Concept Generation Agent (CGA)} explores early-stage solutions through ideation tools and creative reasoning, often invoking supporting agents such as Brainstorming or Empathy agents to stimulate divergent thinking. Existing computational design methods from previous research, such as InnoBID, AutoTRIZ, and data-driven design-by-analogy \cite{Jiang2025autotriz,Zhu2023bio,jiang2022dba}, can be incorporated into the CGA to enhance concept exploration, guide reasoning logic in foundation models, or systematically expand the search space of possible solutions. In engineering design, however, concept generation must balance creativity with physical and operational feasibility. The CGA therefore integrates physical reasoning, such as kinematic compatibility, material limitations, and safety constraints, into its generative exploration to ensure that creative outputs remain grounded in practical design conditions.

\textbf{Embodiment Generation Agent (EGA)} transforms selected concepts into more concrete system architectures, structuring design concepts into coherent assemblies that define the design space and directly constrain downstream possibilities. A well-structured embodiment architecture provides the backbone of the design, determining feasibility and future flexibility. During this process, the EGA may draw on web-search agents or design databases for reference information, as well as creatively engage supporting agents (e.g., Brainstorming agents) to propose alternative system-level arrangements. This stage is therefore critical in bridging abstract concepts with practical engineering design. In engineering design, this translation stage also requires managing cross-domain dependencies, such as thermal-structural coupling or electro-mechanical integration, and coordinating models at different fidelity levels. The EGA thus plays a critical role in bridging abstract concepts with practical multi-physics and multi-scale design representations.

\textbf{Detailed Modeling Agent (DMA)} creates and refines the geometry, specifications, and parameters of the design to generate manufacturable, high-fidelity models. The DMA works closely with supporting agents such as CAD agents and design databases to ensure consistency and feasibility. In current research, CAD generation is being explored through diverse approaches \cite{Zhou2025}, including command sequence modeling (as token generation), programmatic CAD script generation (as code generation), video-based screen learning to mimic human operation, and multimodal contrastive learning. As the field is still rapidly evolving without convergence to a single paradigm, the DMA serves as a flexible interface for integrating these emerging technologies. In engineering applications, a key difficulty lies in encoding geometric intent, tolerance logic, and boundary conditions in a machine-interpretable form. The DMA can serve as a bridge between symbolic design representations and continuous geometric spaces, enabling automated manufacturability checks and iterative refinement.

\textbf{Design Optimization Agent (DOA)} applies generative or simulation-based optimization techniques to improve the design against multiple objectives and constraints. Simulation can involve domain-specific software tools for detailed performance validation, or numerical methods to assess general metrics (e.g., efficiency, robustness, or safety). Design optimization is inherently iterative: the DOA may orchestrate wide loops across CGA, EGA, and DMA for system-level improvements, or small loops within a specific stage to address local refinements. In engineering design, such optimization must integrate both data-driven and physics-based reasoning, often constrained by uncertain boundary conditions and real-world manufacturability limits. The DOA is responsible for planning and executing these hybrid optimization cycles, ultimately delivering a design that balances performance, feasibility, and resource constraints across the full design workflow.

Figure 7 illustrates an exemplar coordination among three agents (CGA, EGA, and DMA). The CGA transforms design requirements into initial conceptual designs by leveraging knowledge databases, design heuristics, and brainstorming agents. The EGA then translates these abstract concepts into structured system architectures using system modeling tools, reference knowledge, and requirement analysis agents to ensure functional feasibility and scalability. After that, the DMA creates manufacturable parametric geometries, performs structural and production-oriented evaluations, and validates compliance with design specifications. This localized internal workflow within the entire multi-agent system bridges early-stage concept development with downstream embodiment engineering design and its detailed modeling.

\begin{figure}[H]
	\centering
	\includegraphics[width=16cm]{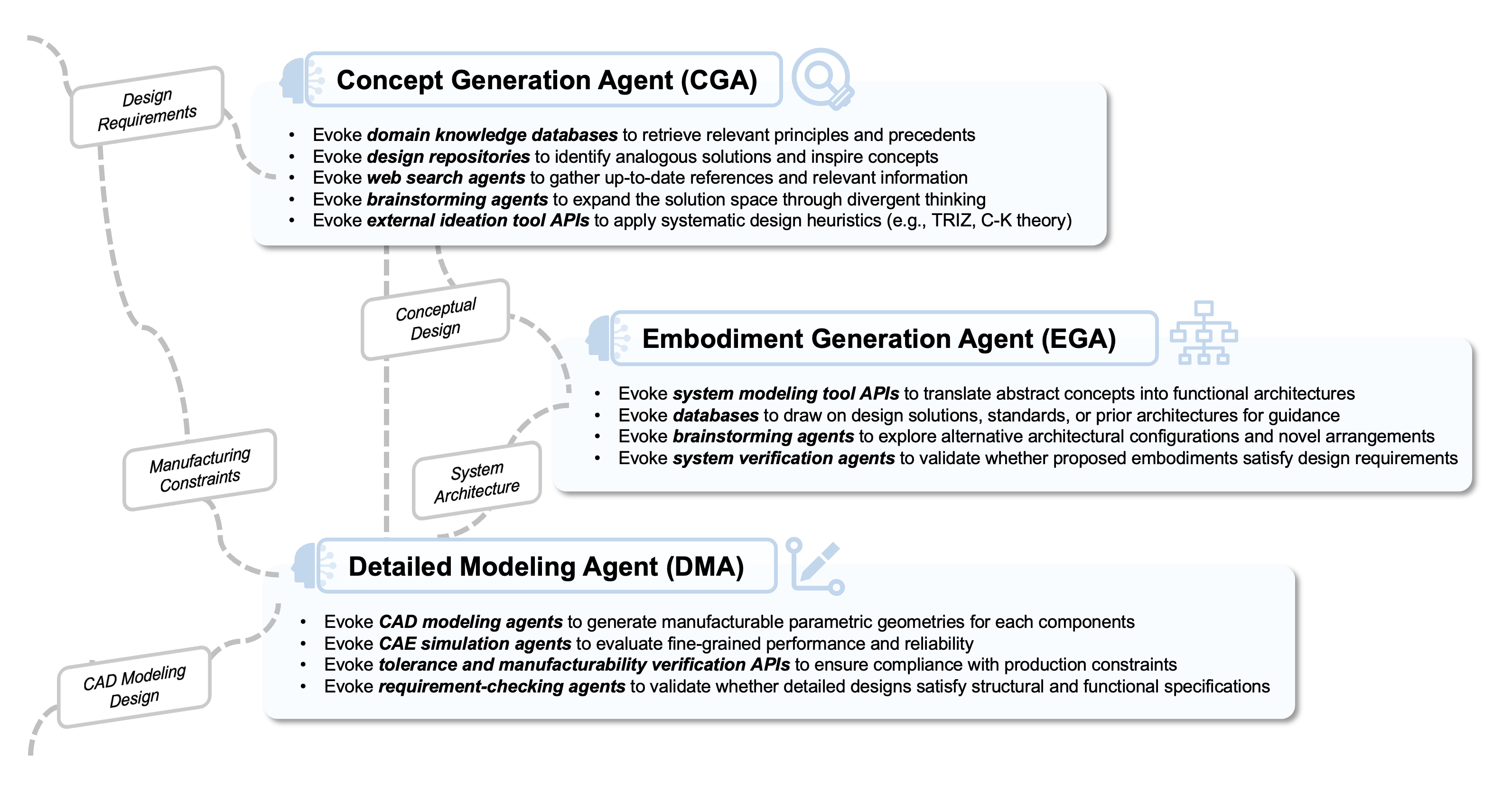}
	\caption{Illustration of Collaboration and Coordination Among Different Design Agents}
	\label{fig:fig7}
\end{figure}

The stage-level design agents are not static modules, but interactive and adaptive components. Each stage-level agent may operate as an individual foundation model-powered agent or as a coordinated group of agents sharing the same objective. Throughout the design process, each agent can dynamically take the lead in decomposing tasks, triggering downstream agents, and coordinating others as needed. To support complex operations, each stage-level agent accesses a shared pool of functional agents that provide domain-general capabilities across the design process, such as web search agent, database agent, and CAD modeling agent. The functional agents are also equipped with capabilities for planning, adaptation, and system information sharing, enabling more intelligent and coordinated collaboration with stage-level agents. While Figure 6 illustrates a representative set of agents, real-world design environments may incorporate additional and domain-specific agents to meet specialized functional needs. The entire system interacts with human designers, who initiate tasks, receive design outputs, and provide feedback to agents at any stage, through a centralized communication interface. Within the system, agents can access external tools and services via standardized protocols \cite{Anthropic2024}. All agents operate within a shared information environment, accessing to both short-term and long-term memory, which support cross-agent coordination and cumulative learning throughout the design process.

Recent studies in engineering design have begun validating multi-agent AI systems in various cases, including aerodynamic vehicle optimization, mechatronic system synthesis, and automated CAD modeling \cite{Wang2025agent,Massoudi2025,Panta2025,Zhang2025gpt,Chen2025cad,Elrefaie2025,Neema2025}. Combined with a design-process-based ontology, the framework proposed in this section could provide a foundation for understanding these emerging developments and guiding future research toward scalable and end-to-end design automation.

\section{Discussion}
\label{sec:sec7}

While recent studies have demonstrated promising early prototypes of multi-agent systems for design, several key challenges must still be overcome before such systems can be effectively applied to complex engineering problems. In this section, we discuss these main challenges and outline potential directions for Intelligent Design 4.0, focusing on three core aspects: (1) the design data foundations required to enhance the capabilities of foundation models and agents, (2) the mechanisms that enable effective collaboration and coordination among multiple agents, and (3) the formulation of increasingly complex and interconnected design problems and objectives.

\subsection{Design Data Foundations}

A fundamental prerequisite for advancing ID 4.0 lies in establishing robust design data foundations. Foundation models obtain emergent capabilities from training on massive and diverse datasets; however, engineering design data differs substantially from the plain text or image typically used in pretraining. Design data is heterogeneous, multi-modal, and often fragmented, presenting unique challenges for applying foundation models in this domain \cite{Ahmed2025}. Engineering design spans a wide range of domains and continuously evolving requirements, making it difficult for knowledge learned from general-purpose datasets to capture the domain-specific principles that guide the design process. Foundation models therefore need to be continuously fine-tuned or adapted using representative and up-to-date design data tailored to specific design objects and contexts. When the volume and diversity of such data are insufficient, foundation models struggle to generalize beyond narrow tasks, and the downstream agents built upon them may exhibit degraded reliability in reasoning, coordination, and decision-making. Establishing robust data foundations is therefore essential for enabling foundation models and multi-agent systems to function effectively in engineering design.

Several challenges highlight the current limitations of design data foundations. First, the scale of available design data is insufficient. Unlike natural language or image data, engineering design datasets are often domain-specific and limited in size, which restricts the ability of foundation models to generalize across tasks. Second, heterogeneity and inconsistency in data representations create significant quality issues. Different design domains adopt distinct modeling standards, ontologies, and representation forms, making it difficult to align data for cross-domain learning and agent collaboration. Third, real-world design constraints are closely linked to upstream and downstream factors such as supply chain availability, material properties, and manufacturability requirements, yet these considerations are rarely reflected in publicly accessible design datasets. Fourth, the absence of iterative data trails is a major limitation. Whereas scientific datasets often capture detailed experimental iterations, design processes typically leave only partial records, making it difficult for agents to learn from the progression of decision-making. Additionally, restricted accessibility remains a significant barrier. High-quality design data is proprietary and siloed within industrial organizations, making it difficult for researchers to access and limiting its potential as a shared infrastructure for intelligent design systems.

For multi-agent ID 4.0 systems, addressing these challenges is critical. Although LLMs have demonstrated broad generalization capabilities in design-related domains and tasks, they remain limited in many concrete design tasks (e.g., complex CAD modeling) and in specific design scenarios. For example, current datasets commonly used in research (e.g., DeepCAD \cite{Wu2021}) are generated through synthetic methods and thus lack the complexity, diversity, and fidelity of real-world engineering models \cite{Zhou2025}. Constructing large-scale and more realistic design datasets will be essential for enhancing the functional capacity of foundation models, thus improving the quality of multi-agent task execution. High-quality design data can also help agents better interpret design objectives and establish reliable standards. For instance, in the interaction between designers and the requirement analysis agent, providing reference designs from related datasets can guide requirement extraction, alignment, and validation.

Researchers and developers can contribute to building such data foundations in several ways: (1) curating domain-specific yet open-access datasets that capture both functional and physical aspects of design; (2) developing unified data schemas and ontologies to standardize representations across domains; (3) integrating lifecycle and supply-chain data streams to contextualize design choices; and (4) exploring effective protocols that allow agents to update and refine shared datasets as design processes continually evolve.

\subsection{Agent Collaboration Mechanisms}

Current research in multi-agent design systems often relies on a rigidly defined workflow to orchestrate collaboration among agents \cite{Wang2025agent}. While such efforts are valuable for prototyping and benchmarking, they may be limited in more complex and open-ended design scenarios, where creative exploration and emergent strategy are essential. To move beyond these limitations, ID 4.0 demands more flexible and dynamic coordination mechanisms, potentially inspired by concepts from self-organizing systems or multi-agent negotiation \cite{Zhu2025bench}. This remains an open question for design practice: How can design agent teams autonomously allocate tasks, exchange knowledge, and evolve their collaboration strategies effectively? In the engineering design process, this challenge is amplified by the inherently multi-disciplinary and interdependent nature of design activities. Agents must coordinate across heterogeneous representations, while ensuring semantic consistency and contextual understanding. Task allocation cannot rely solely on static rules but must adapt dynamically as design requirements evolve or new constraints emerge. Similarly, knowledge exchange among agents demands shared ontologies and intermediate representations that capture geometry, physics, and design intent in machine-interpretable forms. Overcoming these challenges requires advances in ontology alignment, cross-domain reasoning, and negotiation protocols that allow agents to collaboratively decompose, recombine, and refine design tasks in realistic, constraint-rich environments.

In addition, design rarely involves a single objective in practice. Instead, it must reconcile competing goals and non-ideal constraints such as manufacturability, supply chain disruptions, cost, and sustainability requirements \cite{Xiong2019,Gibson2021,Chen2010}. A system that performs optimally in simulation may still fail if it cannot be manufactured efficiently, sourced reliably, or maintained over its lifecycle. To cope with such realities, design agents not only need to optimize across multiple objectives but also to negotiate trade-offs when conflicts arise. Negotiation among different agents over criteria (such as cost versus performance, speed versus sustainability, or reliability versus scalability) provides a mechanism for resolving conflicts and reaching consensus, ensuring that design outputs remain both feasible and value-aligned. Such negotiation may involve explicit bargaining protocols, distributed voting schemes, or the mediation of higher-level coordination agents, thereby allowing multi-agent teams to move beyond rigid optimization toward more adaptive, stakeholder-aware solutions.

Furthermore, as foundation models continue to grow in capacity, the boundary between the multi-agent framework and the single “super-agents” may become blurred \cite{Yao2025}. This raises critical questions: At what point could a single model effectively manage the full stack of an end-to-end design workflow? And how should we balance the generality and power of such models against the modularity, interpretability, and fault isolation offered by multi-agent systems? These questions lie at the heart of the orchestration problem in ID 4.0: what is the optimal architecture for intelligent collaboration, distributed agents with explicit communication and specialization, or unified agents with general-purpose capabilities? Future research and practice should explore hybrid architectural strategies that can dynamically evolve with advancements in model capabilities, balancing the strengths of specialization, negotiation, and generality within the systems.

\subsection{Design Problem Formulations}

As engineering design problems increase in complexity, they may require integrated reasoning across disparate physical scales (e.g., nano to macro) and physics domains (e.g., structural, thermal, fluid, electromagnetic) \cite{DeBorst2008}. Complex products such as aerospace systems, robotics, and biomedical devices often involve tight coupling between microstructural properties, component-level geometries, control strategies, and performance in dynamic environments \cite{Zhao2022,Wang2025codesign,Wang2021}. Traditional methods often treat these domains independently, requiring manual iteration across software tools and disciplinary teams. Within the ID 4.0 paradigm, specialized agents could be orchestrated to operate at different modeling scales, enabling coherent and system-level co-design. For instance, in lightweight aerospace component design, a material-level simulation agent could predict anisotropic mechanical properties arising from composite layup patterns, while a structural optimization agent could use these stress-strain and thermal expansion parameters to adjust topology or thickness distributions. The results could then be iteratively evaluated by a manufacturability agent to ensure the optimized structure can be fabricated within process constraints such as curing temperature and fiber orientation tolerance. The development of such cross-domain agents requires close collaboration among experts from different disciplines to ensure accurate and effective results.

Despite this potential, current foundation models still lack accurate embedded knowledge in fields such as nonlinear mechanics, heat transfer, or optics \cite{Shen2025}. Their reasoning is largely statistical rather than governed by physical laws. In practice, integrating physics engines or world models with LLMs or VLMs is challenging due to mismatched representations between symbolic solvers and token-based models, the computational cost of fine-scale fidelity, and the need to ensure outputs remain consistent with physical laws. Addressing these gaps will require intermediate representations that bridge symbolic and statistical reasoning \cite{Ma2024}, as well as coordinated efforts on expanding design data foundations and effective model architectures. Advancing such hybrid systems could ultimately yield cross-scale design agents capable of reasoning from atomic structures to macro-level artifacts \cite{Agarwal2025}.

In addition, the formulation of design goals themselves is expected to evolve. Engineering design has traditionally been human-centered, with goals specified by users based on their requirements \cite{Zhu2024}. However, as agentic AI systems grow more capable of reasoning and contextual learning, they may begin to formulate goals autonomously. For example, when faced with underspecified prompts, a future ID 4.0 system may extrapolate user intent, identify broader opportunities, or suggest higher-level objectives previously unconsidered. This shift challenges the assumption that humans alone define meaningful goals, and positions agents as partners in ideation capable of autonomous problem discovery \cite{Yang2025}.

At the same time, the shift toward machine-defined goals raises fundamental ethical and legal questions \cite{Mukherjee2025}. Who is accountable for a machine-generated goal? How do we ensure that such goals respect cultural, environmental, or organizational norms? And what does authorship mean when systems contribute not only as problem-solvers but also as problem-setters? To better align ID 4.0 systems with human values, we need alignment on two fronts: technically, through transparent reasoning processes and human-in-the-loop validation; and socially, by embedding ethical norms and stakeholder priorities into system design. More importantly, relevant policies should ensure that system actions are traceable, responsibilities are well-defined, and human intervention remains possible when needed.

\section{Conclusion}

Intelligent Design 4.0 marks a critical inflection point in the evolution of design intelligence, transitioning from data-driven assistance to multi-agent autonomy empowered by foundation models such as LLMs. In this paper, we reviewed the paradigm evolution of intelligent design through four stages, and proposed a framework for implementing ID 4.0. Beyond reviewing past developments, we have outlined future directions and open challenges, including the need for robust design data foundations, more flexible mechanisms for agent collaboration, and new approaches to formulating multi-scale and evolving design problems and goals. We hope this paper could offer both conceptual foundations and actionable perspectives to guide researchers, designers, and developers in advancing the next generation of intelligent design.

\section*{Disclosure statement}

No potential conflict of interest was reported by the author(s).

\bibliographystyle{unsrtnat}
\bibliography{references}

\end{document}